\documentclass[11pt]{article}
\usepackage{moriond,epsfig}

\newcommand{\epem}              {\ensuremath{\mathrm{e^+e^-}}}

\newcommand{\as}                {\ensuremath{\alpha_\mathrm{S}}}
\newcommand{\ash}               {\ensuremath{\hat{\alpha}_\mathrm{S}}}
\newcommand{\mz}                {\ensuremath{m_{\mathrm{Z^0}}}}
\newcommand{\asmz}              {\ensuremath{\as(\mz)}}

\newcommand{\stat}              {\ensuremath{\mathrm{(stat.)}}}

\newcommand{\expt}              {\ensuremath{\mathrm{(exp.)}}}
\newcommand{\had}               {\ensuremath{\mathrm{(had.)}}}
\newcommand{\theo}              {\ensuremath{\mathrm{(theo.)}}}
\newcommand{\ycut}              {\ensuremath{y_{\mathrm{cut}}}}

\newcommand{\chisq}  {\ensuremath{\chi^2}}
\newcommand{\chisqd} {\ensuremath{\chi^2/\mathrm{d.o.f.}}}
\newcommand{\xmu}      {\ensuremath{x_{\mu}}}
\newcommand{\bbbar}             {\ensuremath{\mathrm{b\bar{b}}}}

\begin{document}
\vspace*{4cm}
\title{MEASUREMENT OF THE STRONG COUPLING $\alpha_S$ FROM THE 3-JET
RATE IN \epem\ ANNIHILATION USING JADE DATA }

\author{ S. KLUTH }

\address{Max-Planck-Institut f\"ur Physik, F\"ohringer Ring 6, 80805
Germany}

\maketitle

\abstracts{ We describe a measurement of the strong coupling
  \asmz\ from the 3-jet rate in hadronic final states of
  \epem\ annihilation recorded with the JADE detector at
  centre-of-mass energies of 14 to 44~GeV.  The jets are reconstructed
  with the Durham jet clustering algorithm.  The JADE 3-jet rate data
  are compared with QCD predictions in NNLO combined with resummed
  NNLA calculations.  We find good agreement between the data and the
  prediction and extract
\begin{displaymath}
  \asmz= 0.1199\pm0.0010\stat\pm0.0021\expt\pm0.0054\had\pm0.0007\theo\;\;.
\end{displaymath} }

\section{Introduction}

We report on the first measurement of \asmz\ from the 3-jet rate with
the Durham algorithm with matched NNLO+NLLA QCD
calculations~\cite{schieck12}.  The first measurement of \as\ from
$R_3$ with NNLO QCD calculations was shown in~\cite{dissertori09b}.

\section{JADE Detector and Data}

The JADE detector was a universal and hermetic detector covering a
solid angle of almost $4\pi$.  The interaction point was surrounded by
a large tracking detector (jet chamber) of 1.6~m diameter and 2.4~m
length inside a solenoid magnet coil with a magnetic field of 0.48~T.
Outside of the magnetic coil was the electromagnetic calorimeter
consisting of 2520 lead glass blocks in the barrel section and 96 lead
glass blocks in each endcap with a total acceptance of 90\% of $4\pi$.
The measurement of hadronic final states relies mainly on these two
detector systems.  More details can be found e.g.\ in~\cite{naroska87}.

The data used in the analysis are from the JADE experiment which
operated at the PETRA \epem\ collider at DESY in Hamburg, Germany,
from 1979 to 1986.  The main data samples were collected at
centre-of-mass (cms) energies of 14, 22, 35, 38 and 44~GeV.  The
integrated luminosities range from about 1/pb at 14 and 22~GeV to about
100/pb at 35~GeV and correspond to sample sizes of O($10^3$) events at
14, 22, 38 and 44 GeV and O($10^5$) events at 35 GeV.

\section{QCD Predictions}

The Durham jet clustering algorithm~\cite{durham} defines
$y_{ij}=2\min(E_i,E_j)^2(1-\cos\theta_{ij})/s$ as distance in phase
space between a pair of particles or jets $i$ and $j$ with energies
$E_i,E_j$ and angle $\theta_{ij}$ between them.  The pair with the
smallest $y_{ij}$ is combined by adding their 4-vectors, the particles
or jets $i,j$ are removed and the combined 4-vector is added.  This
procedure is repeated until all $y_{ij}>\ycut$.  The 3-jet rate for a
given value of \ycut\ at a cms energy $Q=\sqrt{s}$ is defined as
$R_3(\ycut,Q)=N_{3-jet}(\ycut,Q)/N(Q)$, where $N_{3-jet}$ is the
number of 3-jet events and $N$ is the total number of events in the
sample.  The 3-jet rate is a measurement of
$\sigma_{3-jet}(\ycut,Q)/\sigma_{had}(Q)$ where
$\sigma_{3-jet}(\ycut,Q)$ is the exclusive 3-jet cross section and
$\sigma_{had}(Q)$ is the total hadronic cross section.

\begin{figure}[htb!]
\begin{tabular}{cc}
\includegraphics[width=0.45\columnwidth]{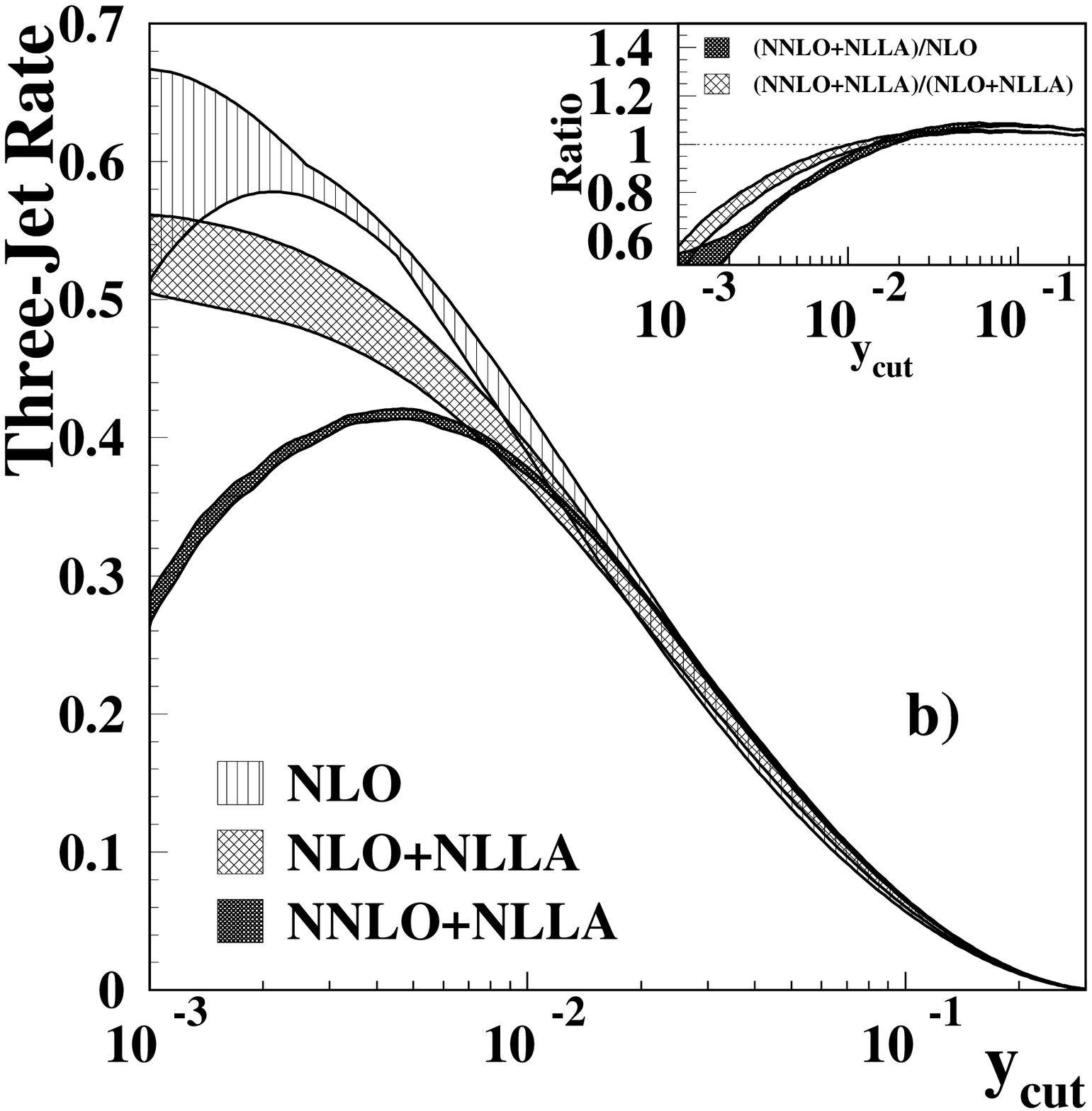} &
\includegraphics[width=0.45\columnwidth]{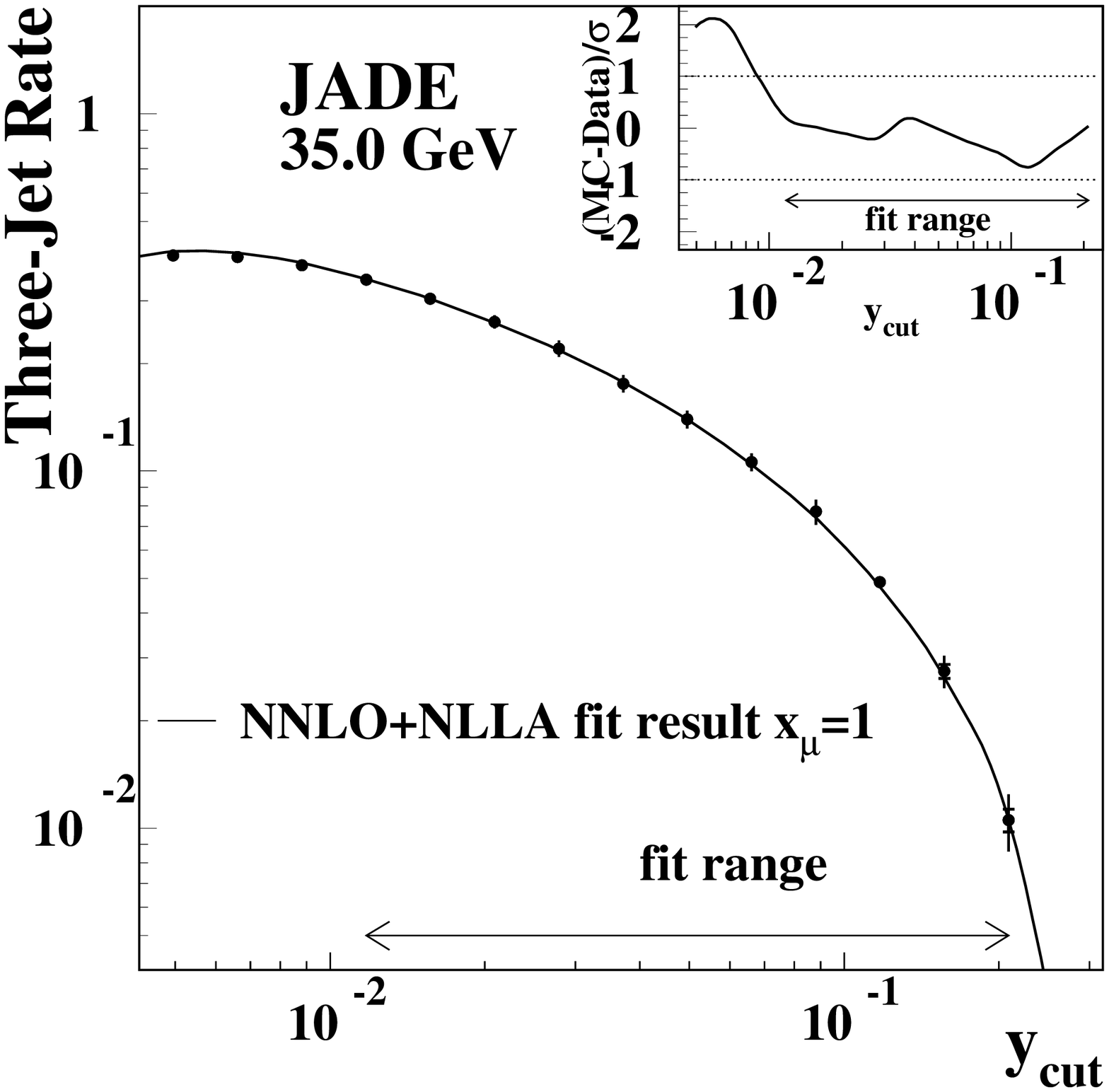} \\
\end{tabular}
\caption[bla]{(left) QCD predictions for $R_3$ in NLO, NLO+NLLA and
  NNLO+NLLA are shown by bands as indicated on the figure.  The widths
  of the bands reflect the renormalisation scale uncertainty.  (right)
  Fit of the NNLO+NLLA prediction to the $R_3$ data at
  $\sqrt{s}=35$~GeV corrected for experimental effects.  The data
  points included in the fit are indicated by the horizontal arrow.
  The insert shows the difference between data and fitted QCD
  prediction divided by the combined statistical and experimental
  error~\cite{schieck12}.}
\label{fig_nlo_fit}
\end{figure}

The NNLO QCD prediction~\cite{gehrmannderidder08,weinzierl09} can be
written as:
\begin{equation}
  R_{3,NNLO}(\ycut,Q)= A(\ycut)\ash(Q) + B(\ycut)\ash^2(Q) + C(\ycut)\ash^3(Q)
\end{equation}
with $\ash(Q)=\as(Q)/(2\pi)$.  The coefficient functions $A(\ycut)$,
$B(\ycut)$ and $C(\ycut)$ are obtained by numerical integration of the
QCD matrix elements in LO, NLO or NNLO.  The resummed NLLA
calculations use an improved resummation scheme~\cite{nagy98c} and are
matched to the NNLO prediction~\cite{schieck12}.
Figure~\ref{fig_nlo_fit} (left) shows these QCD predictions as black
band with renormalisation scale uncertainty defined by multiplying the
renormalisation scale $\mu$ by a factor of $1/2$ or 2.  The other
bands show NLO and NLO+NLLA predictions for comparison.  The
theoretical uncertainties of the NNLO+NLLA prediction are
significantly smaller compared to the less advanced predictions.

\section{Data Analysis}

The data for the 3-jet rate $R_3$ are corrected for the effects of
detector resolution and acceptance and for photon initial state
radiation to the so-called hadron-level using samples of simulated
events.  The expected contributions from $\epem\rightarrow\bbbar$
events are subtracted.  The Monte Carlo generators PYTHIA 5.7, HERWIG
6.2 or ARIADNE 4.11 with parameter settings from OPAL are used to
produce the simulated events together with a full simulation of the
JADE detector.  The corrected data for $R_3$ are well described by the
simulations.

The QCD predictions have to be corrected for effects of the transition
from the partons (quarks and gluons) of the theory to the particles of
the hadronic final state.  These so-called hadronisation corrections
are taken from the samples of simulated events by comparing $R_3$
values after the parton shower has stopped (parton-level) and the
hadron-level consisting of all particles with a lifetime larger than
300~ps.  OPAL has compared for the observable~\footnote{The
  distribution of $y_{ij}$ values for which events change from 2 jets
  to 3 jets.} $y_{23}$ the parton-level predictions of the theory
and the simulation and found agreement within the differences between
the three simulations~\cite{OPALPR431}.  Thus it is justified to use
the simulations to derive the hadronisation corrections, since the
hadronisation systematic uncertainty evaluated by comparing the three
simulations covers any discrepancies.

The theory is compared with the data using a \chisq-fit with
\as\ as a free parameter.  The statistical correlations between
the data points for $R_3(\ycut)$ are taken into account.  Only
data points within a restricted range of \ycut\ are used in the 
fits to ensure that the experimental and hadronisation corrections
are under control and that the QCD predictions are reliable.

Several sources of systematic uncertainty are investigated.
Experimental uncertainties are evaluated by repeating the analysis
with different event selection cuts, reconstruction calibration
versions, corrections for experimental effects, and with different fit
ranges.  The experimental uncertainties are dominated by the different
detector calibrations and the detector corrections based on PYTHIA or
HERWIG.  Hadronisation uncertainties are estimated by changing the
Monte Carlo generator for hadronisation corrections from PYTHIA to
HERWIG or ARIADNE.  The differences between PYHTIA and HERWIG
determine this uncertainty.  Theoretical systematic uncertainties are
found by repeating the fits with the renormalisation scale factor
$\xmu=\mu/Q$ changed from $\xmu=1$ to 0.5 or 2.

\section{Results}

\begin{figure}[htb!]
\begin{tabular}{cc}
\includegraphics[width=0.45\columnwidth]{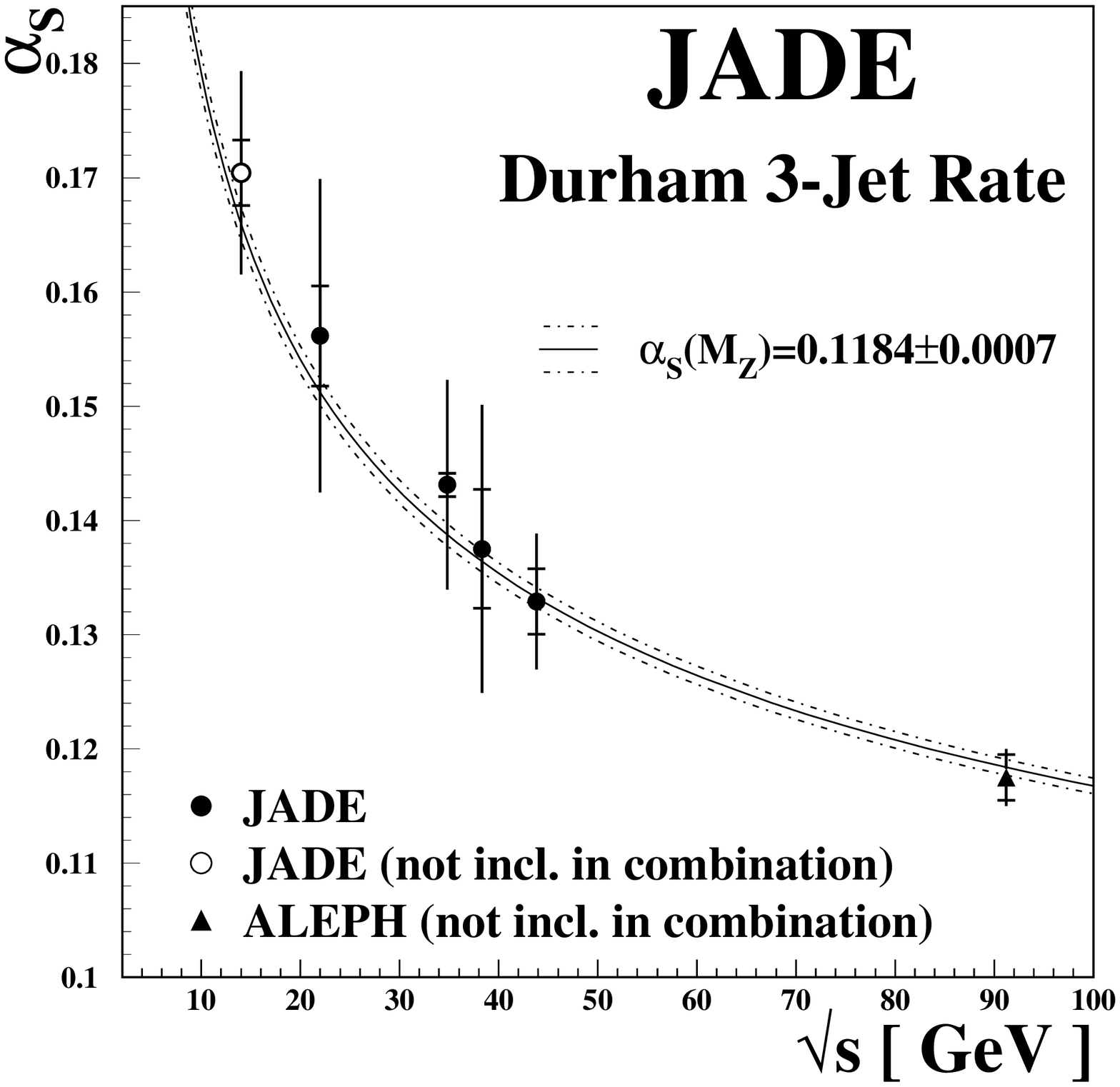} &
\includegraphics[width=0.45\columnwidth]{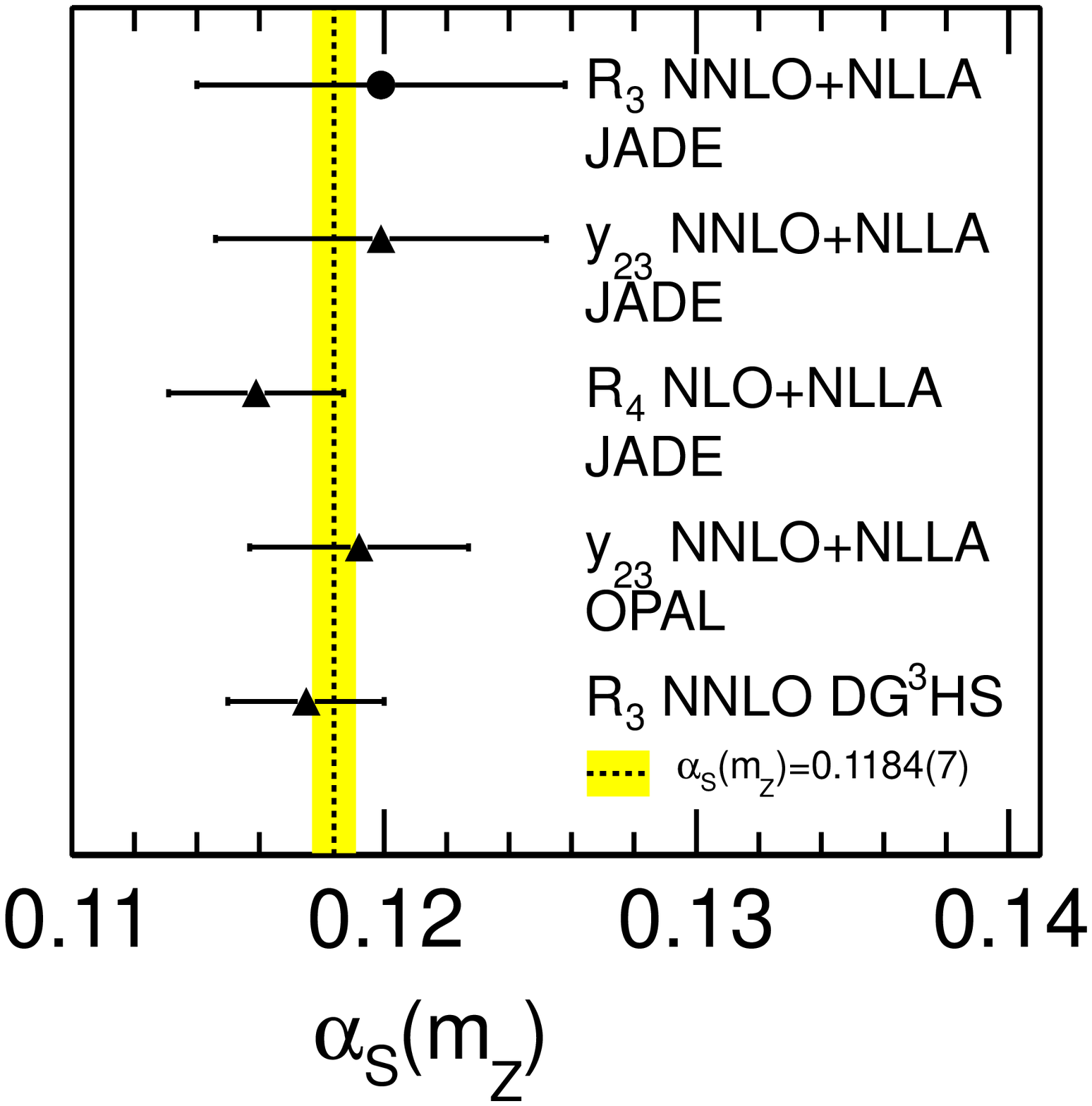} \\
\end{tabular}
\caption[bla]{(left) Results for \as\ from the JADE energy points are
  shown.  The lines give the prediction from the 3-loop QCD evolution
  with uncertainties for the value of \asmz\ as indicated on the
  figure.  (right) The result for \asmz\ from this analysis (solid
  point) is compared with results
  from~\cite{jadennlo,jader4,OPALPR431,dissertori09b} (solid
  triangles) and the current world average
  value~\cite{bethke09,pdg10}. }
\label{fig_fits}
\end{figure}

The fit of the NNLO+NLLA QCD prediction to the 3-jet rate data at
$\sqrt{s}=35$~GeV is shown in figure~\ref{fig_nlo_fit} (right).
The fitted prediction agrees well with the data corrected to the 
hadron-level within the fit range.  The extrapolation to the other
data points also gives a good description of the data.  For this
fit based on statistical errors we find $\chisqd=1.2$.  The fits
at the other cms energies are similar with $1.2<\chisqd<3.8$
except at $\sqrt{s}=14$~GeV where we have $\chisqd=6.3$.  At the
lowest cms energy the hadronisation corrections are significantly
larger compared to the other cms energies.  The individual
fit results for \as\ are shown in figure~\ref{fig_fits} (left)
as a function of the cms energy where they were obtained.

The individual results for \as\ are evolved to \asmz\ using the 3-loop
evolution equations.  Then they are combined into a single value
taking account of correlated experiental, hadronisation and theory
uncertainties as described in~\cite{schieck12}.  The result from
$\sqrt{s}=14$~GeV is excluded from the combined value since it has a
much larger value of \chisqd\ and larger hadronisation corrections
compared to the other results.  The combined value is
\begin{equation} 
  \asmz= 0.1199\pm0.0010\stat\pm0.0021\expt\pm0.0054\had\pm0.0007\theo\;\;.
\end{equation}
The errors are dominated by the hadronisation correction
uncertainties.

As a cross check the analysis is repeated with NNLO QCD predictions
using the same fit ranges with $\xmu=1$.  We find larger values of
\chisqd, a less satisfactory description of the $R_3$ data and larger
uncertainties from variations of the fit ranges compared to the
NNLO+NLLA fits.  The NNLO predictions do not reproduce the slope of
the $R_3(\ycut)$ data as well as the NNLO+NLLA predictions.  A similar
observation can be made in the analysis of~\cite{dissertori09b}.

In figure~\ref{fig_fits} (right) the result of this analysis is
compared with other measurements of \asmz\ using the 3-jet or 4-jet
rate based on the Durham algorithm.  The JADE measurement with
$y_{23}$ is highly correlated with our measurement using $R_3$ and the
good agreement of the results is a strong consistency check.  The
agreement with the other results and with the world average value is
also satisfactory within the uncertainties.

\section{Conclusion}

We have shown the first measurement of \asmz\ using the 3-jet rate
with the Durham algorithm and matched NNLO+NLLA QCD calculations and
data from the JADE experiment.  The agreement between data and the
NNLO+NLLA QCD prediction is improved compared to less advanced
predictions.  The errors are dominated by the hadronisation correction
uncertainties as expected at the low cms energies of the JADE
experiment.  However, the data of the JADE experiment at comparatively
small cms energies can now be analysed with rather good precision
thanks to the progress in perturbative QCD calculations and Monte
Carlo simulations made since the data were recorded.  Our
analysis provides an independent and strong cross check on those
recent QCD calculations made for the LHC which have related
Feynman diagrams or share calculation techniques.

\section*{References}


\end{document}